\documentclass[onecolumn,showpacs,pra,aps]{revtex4}
\usepackage{amssymb}
\usepackage{mathrsfs}
\usepackage{amssymb}
\usepackage{amsmath}
\usepackage{graphicx,graphics}

\begin{document}

\preprint{APS/123-QED}
\title{Diverse classical walking of a single atom in an amplitude-modulated standing wave lattice}
\author{Lin Zhang}
\email{zhanglincn@snnu.edu.cn} \author{H. Y. Kong}
\author{S. X. Qu}
\affiliation{Institute of Theoretical and Computational Physics,
Shaanxi Normal University, Xi'an 710062, P. R. China}
\date{\today}

\begin{abstract}
The classical walking behaviors of a single atom in an
amplitude-modulated standing wave lattice beyond the internal
dynamics are investigated. Based on a simple effective model, we
identify a diversity of dynamic regimes of atomic motion by
periodically adjusting the lattice depth. Harmonic oscillation or
pendulum rotation with classical step-jumping, random scattering
walking, chaotic transportation, quasi-periodic trapped motion and
roughly ballistic free flying are found in this simple model within
different parametric regions by approximate analyses. Our study
demonstrates a complex motion of single atom in modulating optical
lattice beyond the quantum description.
\end{abstract}

 \maketitle 

\section{Introduction}
\label{sec1} %
Nowadays, the fabrication technology steps more and more towards the
atomic scale, and detection on one single molecular or atom is now
an promising technique \cite{Teper,Stibor}. Controlling one atom in
the cavity quantum electrodynamics has now nearly met this step
\cite{Ye,Kuhr,Wilk}. Based on this situation, the motion of one
single atom in a controlled 1-dimensional (1D) optical lattice
becomes a hot topic with a continuous interesting for its basic
theoretical value and reliable applications in quantum information
processing \cite{Monroe}, atom lithography \cite{Arun}, solid state
physics \cite{Niu}, quantum walking algorithm \cite{Karski,Joo},
plasmonic application \cite{Nie,Chang}, quantum-to-classical
transition \cite{Moore} and molecular dynamics. In the optical
lattice, the atom can be accelerated \cite{Doherty} or decelerated
\cite{Horak} by an optical force for the purpose of precise control.
However, in atomic scale, quantum effect starts to play a tricky
role in the atomic dynamics. For example, under a deterministic
linear force, the atomic motion is still beyond definite prediction
in quantum theory because of an intrinsic uncertain variation. That
the quantum motion distinguished from a classical trajectory motion
is actually due to the superposition of the atomic wave packet which
induces interference effect such as dynamical localization
\cite{Zoller}. But, just for one single atom, the external wave
superposition will be excluded \cite{1} and the motion at different
time, also, will not interfere unless a long memory imprinted on the
optical field can feed back to the atomic motion along the classical
trajectory. However this memory will also be cut down by a definite
external optical control. Therefore the quantum motion and the
classical motion meet together at this single atomic scale.

According to quantum descriptions by including internal level
transitions, many kinds of motional behaviors were found in this
single atomic system, such as nonclassical motional states,
ballistic transport, harmonic oscillations, random walks, L\'{e}vy
flights, and chaotic transport
\cite{Meekhof,Prants1,Prants2,Prants4,Prants5,Prants}. Which of
these behaviors are derived from quantum description and which comes
from classical dynamics are still a mixing problem in this system.
This multiple dynamics inspires us to find if there exists a simple
classical model without internal dynamics to reproduce most of the
above dynamical processes. The interesting thing is that, in this
paper, we do find a simple classical model which can produce most of
above behaviors without directly considering internal dynamics. We
identify a rich dynamic picture of the atomic motion in this
classical model in order to pick up all the classical information
from a mixture of description. Our study is not to give differences
between quantum dynamics and classical dynamics because different
descriptions will definitely give different results such as using
different phase-space distribution functions \cite{Wisniacki}. The
only thing to do, in this paper, is to find out all the classical
dynamics of atomic walking within a classical framework and make a
try to show whether some nonclassical motions are more a classical
collective behavior (or a long time statistical behavior) than a
quantum wave description of an individual particle. In order to give
a full picture of the classical motion for a quantum system easy for
check, we closely investigate the classical walking of a single atom
in a controlled optical lattice. We first derive the simple
classical dynamic model in section \ref{sec2}, and find out a
variety of atomic walking behaviors of the model in section
\ref{sec3}. Some approximate analytical solutions on specific
conditions are given to understand the complicated behaviors of
atomic walking in this unsolvable system. Section \ref{sec4}
provides the simple discussions and the main conclusions.

\section{The simple classical model}
\label{sec2}%
What we start with is a widely verified quantum model which
describes the dipole interaction between an atom and a classical
field named Rabi model \cite{Rabi}. The dynamic of a two-level
atom with a dipole moment $\mathbf{d}$ interacting with a
quasi-monochromatic plane-wave field,
\begin{equation}
\mathbf{E}\left(\mathbf{r},t\right)=\mathbf{e}_{\lambda
}\mathscr{E}\left( \mathbf{r},t\right) \cos \left( \nu t-
\mathbf{k}\cdot\mathbf{r} \right), \label{field0}
\end{equation}%
is described by a Hamiltonian of%
\begin{equation}
\hat{H}=\frac{\mathbf{\hat{p}}^{2}}{2m}+\frac{1}{2}%
\hbar \omega _{0}\hat{\sigma}_{z}-\hbar \Omega\left(
\mathbf{r},t\right) \cos \left( \nu t- \mathbf{k}\cdot\mathbf{r}
\right) \left( \hat{\sigma}^{-} \mathbf{+}\hat{\sigma}^{+}\right),
\label{H}
\end{equation}%
where the Rabi frequency is defined by
$\Omega(\mathbf{r},t)=\mathscr{E}\left( \mathbf{r},t\right) \cdot
\mathbf{d}/\hbar$. In Eq.(\ref{field0}), $\mathscr{E}\left(
\mathbf{r},t\right)$ is the temporal envelop of the wave field which
can be definitely controlled in the experiment. The modulated Rabi
frequency $\Omega(\mathbf{r},t)$ expresses the coupling intensity of
the field mode with the atomic dipole $\mathbf{d}$, and
$\hat{\sigma}^{+}$ ($\hat{\sigma}^{-}$) is the atomic level raising
(lowering) operator. Hamiltonian (\ref{H}) is a widely used model to
study the interaction between a two-level atom and a classical field
with different controlling modes denoted by $\Omega(\mathbf{r},t)$.

However, what we consider here is a simple case: a single atom
interacts with a standing wave mode in a microcavity with only its
axial direction, $x$, getting involved. In this case, the
Hamiltonian becomes
\begin{equation}
\hat{H}=\frac{\hat{p}^{2}}{2m}+\frac{1}{2}\hbar \omega
_{0}\hat{\sigma}_{z}-\hbar \Omega \left( x,t\right) \cos \nu t\left(
\hat{\sigma}^{-}\mathbf{ +}\hat{\sigma}^{+}\right) , \label{III-2}
\end{equation}%
where $\Omega \left( x,t\right) $ depends on the axial cavity-mode
with an envelop modulated by  the input field. In a resonant mode
field, Eq.(\ref{III-2}) reduces to a so-called double resonance
model \cite{Kolovsky} with an effective Hamiltonian of (see Appendix
A for details)
\begin{equation}
\hat{H}=\frac{\hat{p}^{2}}{2m}-\hbar \lambda \cos (\chi t)\cos
\left( k \hat{x}\right), \label{DRM-1}
\end{equation}%
where the key parameter $\chi$ is the modulation frequency of the
field amplitude. Eq.(\ref{DRM-1}) is an extensively considered model
\cite{Saif,Pritchard,Steck1,Steck2} along with the phase-modulated
model \cite{Raizen1} in quantum chaos. Here the Hamiltonian
describes the dynamics of an atom (or a polar molecule) moving in an
amplitude-modulated standing wave under an effective coupling
constant $\lambda$. However in quantum theory, this model is
time-dependent and difficult to be solved in a normal technique
\cite{Korsch}. For one single atom, we can introduce the following
dimensionless classical variables $H= \langle \hat{H}\rangle/\hbar
\omega _{r}$, $p = \langle \hat{p}\rangle/\hbar k$, $x=k
\langle\hat{x}\rangle $, and dimensionless parameters $\lambda
^{\prime }=\lambda /\omega _{r}$, $\omega =\chi /\omega _{r}$,
$t^{\prime }=\omega _{r}t$, scaled by recoil frequency of a single
photon $\omega _{r} =\hbar k^{2}/m$. The label $\langle \hat{O}
\rangle=O$ means the expectation value of the corresponding quantum
operator $\hat{O}$ and this is a very good approximation for one
single atom. In atomic scale, a single atom can be treated as a
classical point with its position identified by the center of mass,
and no fluctuations induced by spatial superposition exist here.
Therefore the classical model can be obtained (primes are omitted)
by
\begin{equation}
H(t)=\frac{p^{2}}{2}-\lambda \cos x \cos \omega t. \label{DRM}
\end{equation}%
According to Eq.(\ref{DRM}), the
classical dynamic equations are%
\begin{eqnarray}
\dot{x} &=&\frac{\partial H}{\partial p}=p, \quad \dot{p}
=-\frac{\partial H}{\partial x}=-\lambda \sin x\cos \omega t,
\label{Dyeqution}
\end{eqnarray}%
or%
\begin{equation}
\ddot{x}+\lambda \cos \omega t\sin x=0.  \label{Dyeqution1}
\end{equation}%
Eq.(\ref{Dyeqution1}) looks simple, but it is a time-dependent
nonlinear differential equation and no closed analytical solution
for an arbitrary parameter is available \cite{Reichl0}.

The steady solutions of Eq.(\ref{Dyeqution1}) are clear at the
antinode sites for $x^{*}=0, \pm \pi, \pm 2\pi, \cdots$ and
$p^{*}=0$, but they are not always stable because the trapping
potential is time-dependent in the way of
\begin{eqnarray*}
V\left( x,t\right) &=&-\lambda \cos\left(\omega t\right) \cos x
=-\frac{\lambda }{2}\left[ \cos \left( x+\omega t\right) +\cos
\left( x-\omega t\right) \right],
\end{eqnarray*}%
which acts on the atom as a superposition of two potential waves
propagating in opposite directions. In a high fineness cavity, this
potential is usually used to control atomic motion and can be easily
generated by a modulation of standing cavity-mode.
\begin{figure}[tbp]
\begin{center}
\includegraphics[width=230pt]{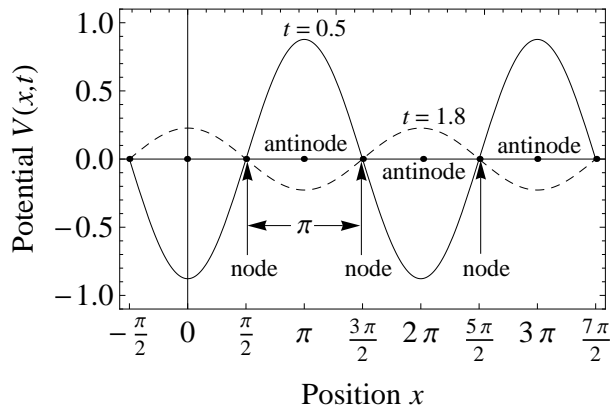}
\end{center}
\caption{The effective potential along lattice $x$ at different time
$t=0.5$ (solid line) and $t=1.8$ (dashed line) with parameters
$\lambda=1,\omega=1$. The time $t$ is scaled by recoil frequency
$\omega_r$ and the position $x$ is scaled by the wavelength of the cavity mode.}
\label{Figure1}
\end{figure}
Fig.\ref{Figure1} shows the varying potentials at two different
times along the axial direction. The nodes (indicated by arrows) or
antinodes of the potential lattice are fixed with a spatial period
of $\pi$ but the amplitude of the lattice wave changes with time
with a period of $t=2\pi / \omega$. The amplitude $\lambda$ at
antinodes denotes the atom-lattice coupling indicating a trapping
ability of the atom at antinode, so we can call it as a lattice
depth. All the above lattice parameters can influence the dynamic of
the atom and all can be controlled by an external optical field.

However, due to atomic motion, the potential lattice felt by
the atom also depends on the momentum of the atom. If we suppose
that the velocity of the atom is a slowly varying quantity during a
time period of $[0,t]$, we can set
\begin{equation*}
x(t)=\int_{0}^{t} p(\tau)d\tau=\langle p\rangle\cdot t \approx p_0
\cdot t,
\end{equation*}%
where $\langle p\rangle$ is a time average of momentum which can be
replaced by its initial value of $p_0$. In this case, the optical
force will be
\begin{equation}
F(t)=-\frac{\partial V}{\partial x}=-\frac{\lambda }{2}\left[ \sin
\left( \omega +p_0\right) t-\sin \left( \omega -p_0\right) t\right]
,\label{force}
\end{equation}%
which clearly indicates that there are two driven forces exerted on the
atom with two different frequencies depending on atomic momentum,
suggesting two nonlinear resonance points of the dynamics at $\omega
\pm p_0$ \cite{Kolovsky}. Therefore, the motion of the atom across
the varying lattice field is dependent on two aspects: the lattice
field and the state of the atom. Although this model described by Eq.(\ref{DRM}) is
simple, it can display most of the dynamic
behaviors found in
\cite{Prants1,Prants2,Prants4,Prants5,Prants} without directly
including the internal dynamics.

\section{Diversity of walking behavior}
\label{sec3}%
\subsection{Oscillation with random step-jumping}
\begin{figure}[htp]
\begin{center}
\includegraphics[width=240pt]{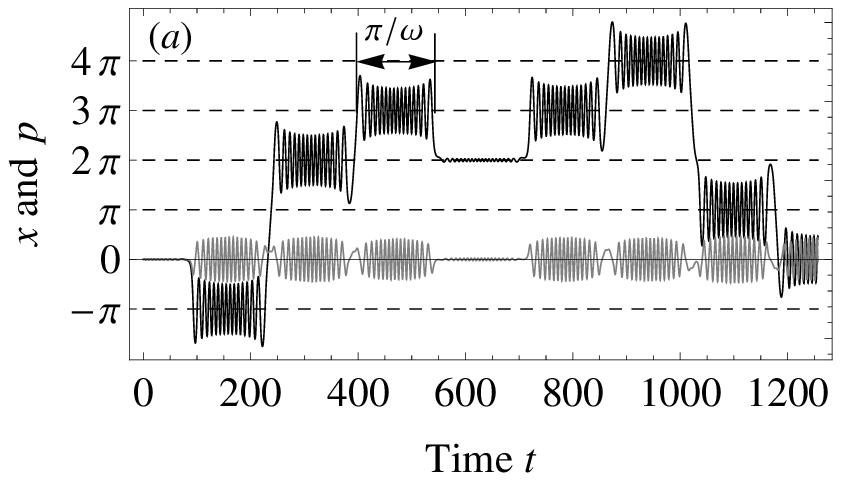}
\includegraphics[width=240pt]{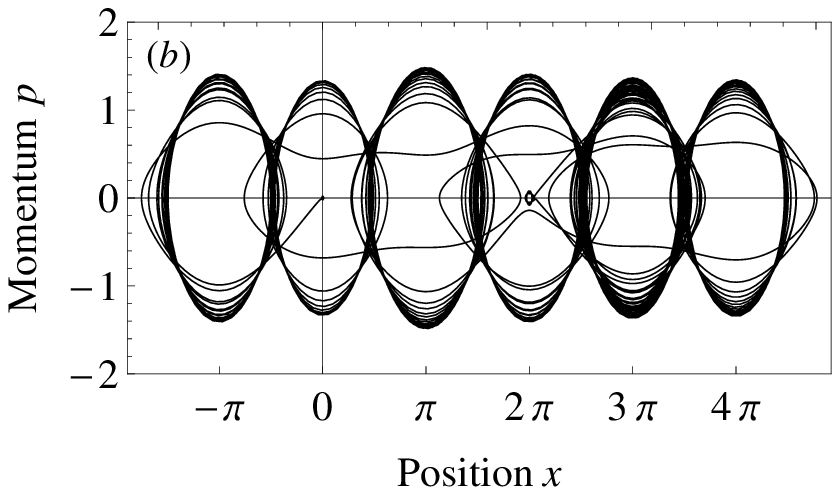}
\end{center}
\caption{A typical walking trajectory of a cold atom in slowly
varying standing wave with $\lambda =1$, $\omega=0.02$, the initial
position and momentum $x_{0}=0$, $p_{0}=0.02 $. (\textit{a}) The
evolution of the atomic position (black line) and momentum (gray
line). (\textit{b}) The corresponding orbit in phase space.}
\label{figure2}
\end{figure}

Under the condition that the amplitude of the lattice is varying
slowly, the cold atom ($p< 2\sqrt{\lambda}$) will be trapped at the
bottom of the lattice potential for a long time, oscillating around
the antinodes at $\sin x^{*}=0$, i.e. $x^{*}=\pm n \pi,
n=0,1,2\cdots$. In Fig.\ref{figure2}, a typical walking process of
this case is shown by a numerical simulation on
Eq.(\ref{Dyeqution}). Fig.\ref{figure2}(a) displays a temporal
position (black line) and momentum (gray line) of the atomic walking
and Fig.\ref{figure2}(b) is its orbit in the phase space.
We can see that the atom first oscillates around one of antinodes
(indicated by the horizonal dashed lines) and then jumps randomly in
an integer steps of $\pi$ to a left or right antinode after a time
interval of $\pi/\omega$. The random atomic jumping to a new site is
due to the loss of stability of the former site when the coupling
intensity $\lambda(t)=\lambda \cos(\omega t)$ turns from positive to
negative. This classical jumping behavior demonstrated in this model
will reversely affect the lattice field and can be traced by the
transmission field from the cavity due to a motion-dependent
detuning effect \cite{Rempe,Kimble}. Although a similar dynamic
behavior was revealed by Domokos and Ritsch \cite{Ritsch}, the
mechanisms of two systems are different. Our system is only for one
single atom and no random Langevin-type noise is necessary during
the dynamics. In order to understand more about this classical
trapping and jumping behavior, we can analyze it by the following
extreme approximations.

\subsubsection{Harmonic Oscillation with random jumping}

For the small modulation frequency of $\omega \ll 1$, the
approximation of $\cos \omega t\sim 1$ is valid for a short time.
The linearization for a cold atom ($p \ll \sqrt{\lambda}$ for
tightly trapping) around the stable positions of $\sin x^{*}=0$,
such as $\sin x\sim x$ around $0$ (for other site we can set
$x^{\prime}=x-x^{*}$) can be used here. Then Eq.(\ref{Dyeqution1})
will reduce to
\begin{eqnarray}
\ddot{x}+\lambda x \approx 0, \quad \omega \ll 1, \label{harmonic}
\end{eqnarray}%
where we set $\lambda >0$ and this enables Eq.(\ref{harmonic}) a
harmonic oscillation solution of
\begin{equation}
x\left( t\right) \approx x\left( 0\right) \cos \left( \sqrt{\lambda
}t\right) + \frac{p\left( 0\right) }{\sqrt{\lambda }}\sin \left(
\sqrt{\lambda } t\right) ,\label{sol1}
\end{equation}%
which describes a main properties of atomic walking in this case.
Eq.(\ref{sol1}) indicates that the oscillation frequency in
Fig.\ref{figure2}(a) is determined by the depth of the trap,
$\lambda$, and gives an important characteristic time of oscillation
period, $2\pi/\sqrt{\lambda }$. As the temporal depth
$\lambda(t)=\lambda \cos \omega t$ changes slowly with time, the
oscillation frequency will adiabatically follow with
$\sqrt{\lambda(t)}$. When $\lambda(t)$ becomes negative, the
oscillation frequency $\sqrt{\lambda(t)}\rightarrow
i\sqrt{|\lambda(t)|}$, and the atom will escape exponentially from
the former site in a way of
\begin{equation}
x\left( t\right) \approx x\left( 0\right) \cosh \left( \sqrt{\lambda
}t\right) + \frac{p\left( 0\right) }{\sqrt{\lambda }}\sinh
\left(\sqrt{\lambda}t\right).\label{sol2}
\end{equation}%
Therefore the atom will conduct an oscillation followed by an
escaping jump from the former site.
\begin{figure}[htp]
\begin{center}
\includegraphics[width=220pt]{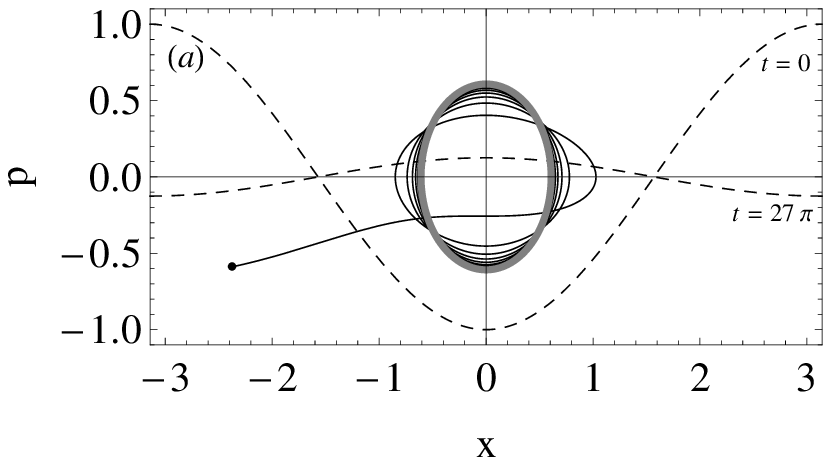}
\includegraphics[width=220pt]{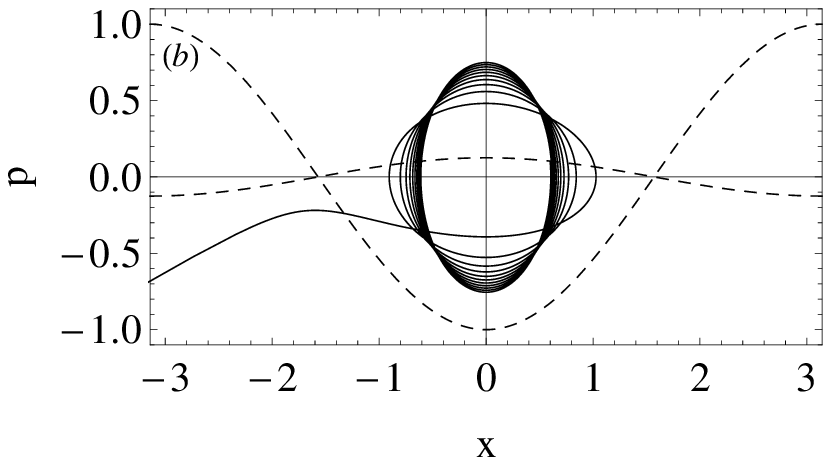}
\end{center}
\caption{The enlarged orbit of a cold atom in phase space with
$\omega=0.02$ and $\lambda =1$. The initial position and momentum
are $x_{0}=0.6$, $p_{0}=0.1 $. (a) The orbit of the rigid simulation
(solid black line) with harmonic approximation (gray thick line);
(b) The approximation orbit of Airy solution of Eq.(\ref{Airy}).}
\label{figure3}
\end{figure}
In order to see the details of this motion, Fig.\ref{figure3}(a)
exhibits a zoomed in orbit around one antinode in phase space. The
black line is the real orbit and the thick gray ellipse stands for
the harmonic oscillation of Eq.(\ref{sol1}), where the potential
lattices at time $t=0$ and at the end of one oscillation $t=27\pi$
are depicted by the dashed lines for reference. As a decrease of
$\lambda(t)$, the elliptic orbit in the phase space expands along
$x$ until $\lambda(t)$ decreases to a value that the atom can pass
through it to another site at a location determined by the escaping
momentum. Therefore the trapping time around antinodes for the cold
atom is about $\pi/\omega$ as indicated in Fig.\ref{figure2}(a),
which is estimated by the time when $\lambda(t)$ changes from
positive value to negative value. In Fig.\ref{figure3}(a), the
oscillating time of the atom around the starting site is about
$t=\pi/2 \omega=25 \pi$, which is just the time taking by the trap
depth $\lambda(t)$ decreasing from maximum to zero.

Certainly, for a more rigorous approximation in above case of
$\omega t \ll 1$, the time dependent part will be $\cos(\omega
t)\approx 1-(\omega t)^2/2$, and the atomic motion can be better
described by a parabolic cylinder function satisfying differential
equation of
\begin{eqnarray}
\ddot{x}+ \lambda(1-\frac{\omega^2 t^2}{2}) x = 0,
\end{eqnarray}%
or even by an Airy function determined by
\begin{eqnarray}
\ddot{x}+ \lambda(\frac{\pi}{2}-\omega t) x = 0,\label{Airy}
\end{eqnarray}%
where the time approximation $\cos(\omega t)=\sin(\pi/2-\omega
t)\approx \pi/2-\omega t$ is used. Fig.\ref{figure3}(b) demonstrates
a better orbit of the Airy solution than a harmonic oscillation at a
short time. But for a longer time the dynamic will dramatically
deviate from the real orbit because of a divergence of Airy
function.

\subsubsection{Pendulum rotation with random step-shifting}
When varying frequency $\omega$ increases near to the oscillation
frequency $\omega \sim \sqrt{\lambda}/2$, the trapping time
$\pi/\omega$ of the cold atom will decrease and the atom will
quickly shift from one site to another like a pendulum swinging or
rotating around different equilibrium sites just for a few periods.
As the motion of atom covers a large range of $x$ relative to stable
antinodes in this case, the spatial linearized Eq.(\ref{harmonic})
will be invalid. Therefore Eq.(\ref{Dyeqution1}) should be
\begin{equation*}
\ddot{x}+\lambda \sin x=0,
\end{equation*}%
which is the well-known pendulum equation with $\lambda>0$.
\begin{figure}[htp]
\begin{center}
\includegraphics[width=228pt]{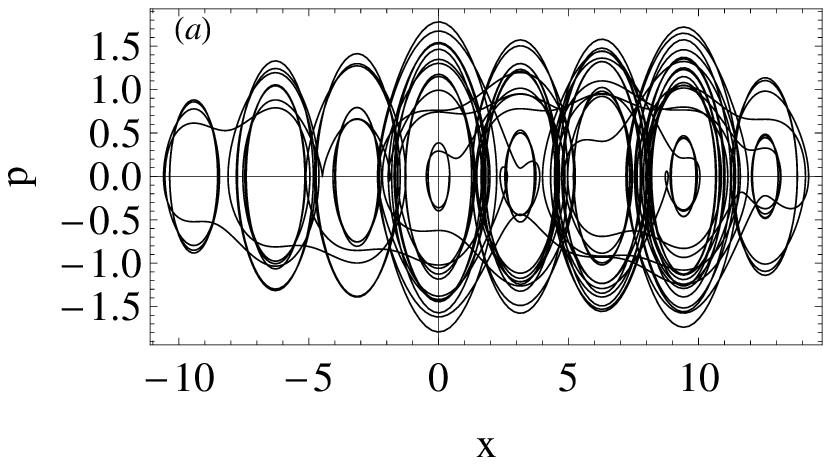}
\includegraphics[width=220pt]{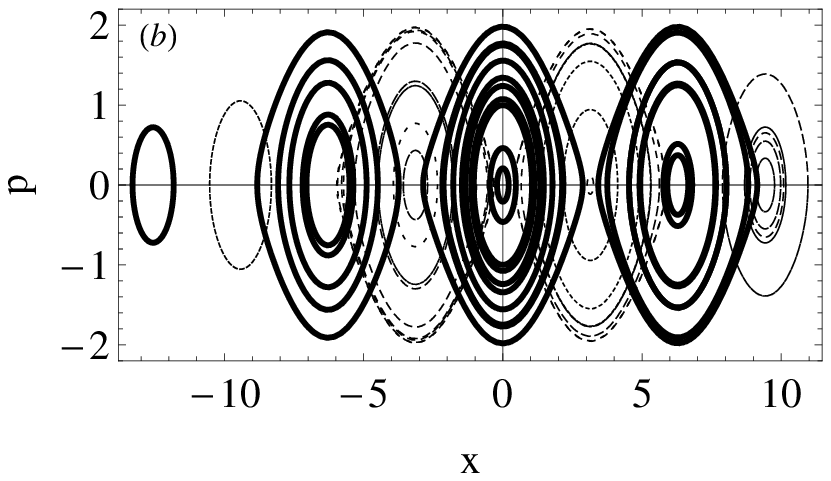}
\end{center}
\caption{The phase space picture of a cold atom walking in a slowly
varying field with $\lambda =1$, $\omega=0.1$ and the initial atomic
momentum $p(0)=0$. (a) The rigorous solution for initial position being
$x(0)=0.4$; (b) The pendulum solution of Eq.(\ref{Pedu-s}) by using
a random step-shifting (step unit $\pi$) position of Gaussian
distribution with mean value of $0.4$ and variance of $\pi/2$.}
\label{figure4}
\end{figure}
Actually, this situation corresponds to an atom moving in a
stationary one-dimensional lattice. When $\lambda
>0$, the integral solution of above equation takes the form of
\begin{equation}
x\left( t\right)= 2\arcsin \left[ \sin \frac{x_{0}}{2}SN\left(
\sqrt{ \lambda }t+K\left( \sin \frac{x_{0}}{2}\right) ,\sin
\frac{x_{0}}{2}\right) \right], \label{Pedu-s}
\end{equation}%
with the initial conditions of $ x\left( 0\right) =x_{0},
\dot{x}\left( 0\right) =0$. The label $SN(\cdot,\cdot)$ is the
Jacobi elliptic function and $K(\cdot)$ is the complete elliptic
integral of the first kind defined by the elliptic integral of
\begin{equation*}
K(k)=\int_{0}^{\pi/2 }\frac{d\theta}{\sqrt{1-k^{2}\sin ^{2}
\theta}}.
\end{equation*}%

Walking orbits in the optical lattice with a higher modulation
frequency $\omega=0.1$ are demonstrated in Fig.\ref{figure4}.
Fig.\ref{figure4}(a) is a rigid simulation and Fig.\ref{figure4}(b)
presents a pendulum samples of phase picture with random initial
positions. We can see the long-time atomic walking can be somehow
(not exactly) explained by the pendulum solutions with shifting
equilibrium positions. As effective $\lambda(t)$ is actually
time-dependent, the length of the pendulum will change gradually
from positive to negative. When $\lambda(t) < 0$, the stable site at
$x^{*}=0,\pm 2\pi,\pm 4\pi,\cdots$ for $\lambda(t) > 0$ will become
unstable and shift to a new stable site at $x^{*}=\pm \pi,\pm
3\pi,\cdots$, which is clearly shown in Fig.\ref{figure2}(a) with a
sequence of $-\pi,2\pi,3\pi,2\pi,3\pi,4\pi,\pi,0\pi,\cdots$. In
Fig.\ref{figure4}(b) the thick black lines are pendulum solutions of
Eq.(\ref{Pedu-s}) for $\lambda
> 0$ and the thin dashed lines are for $\lambda < 0$. We use a
Gaussian random number to pick up initial positions for the pendulum
solutions and the results reveal that a long time atomic walking in
Fig.\ref{figure4}(a) can be illustrated by Eq.(\ref{Pedu-s}) with a
random shifting of equilibrium position of pendulum rotation. The
simulation indicates that this behavior is a combination of pendulum
oscillation with a random shifting of the optical lattice.

According to above analysis, we can conclude that, in a slowly
modulating optical lattice, the cold atom will feel a gradually
changing wave trap, resulting in a harmonic or pendulum rotation
followed by a random step-shifting. In this process, two
characteristic times are important, trapping time $\pi/\omega$ and
oscillation period $2\pi/\sqrt{\lambda}$. In addition, the
step-shifting between antinodes of the lattice has a random
characteristic and can not be predicted because the jumping process
is very sensitive to position and momentum. In this respect, the
atomic walking in modulated standing wave is very similar to
one-dimensional billiard problem that a minor deviation of position
(or momentum) will leads to a dramatically different walking orbit.
However, above conclusion is on the conditions that the optical
lattice is slowly modulated and the atom is very cold. In the
following parts we will find some other specific walking behaviors
beyond these constraints.

\subsection{Random walking and Chaotic transportation}

When the modulation frequency of the lattice is further increased,
the trapping time of the cold atom, $\pi/\omega$, will decrease to
an extent that it is comparable to the shifting time between
different sites, $\pi/\langle p\rangle$ ($\langle p\rangle$ is
average momentum during a walking interval).
\begin{figure}[htp]
\begin{center}
\includegraphics[width=230pt]{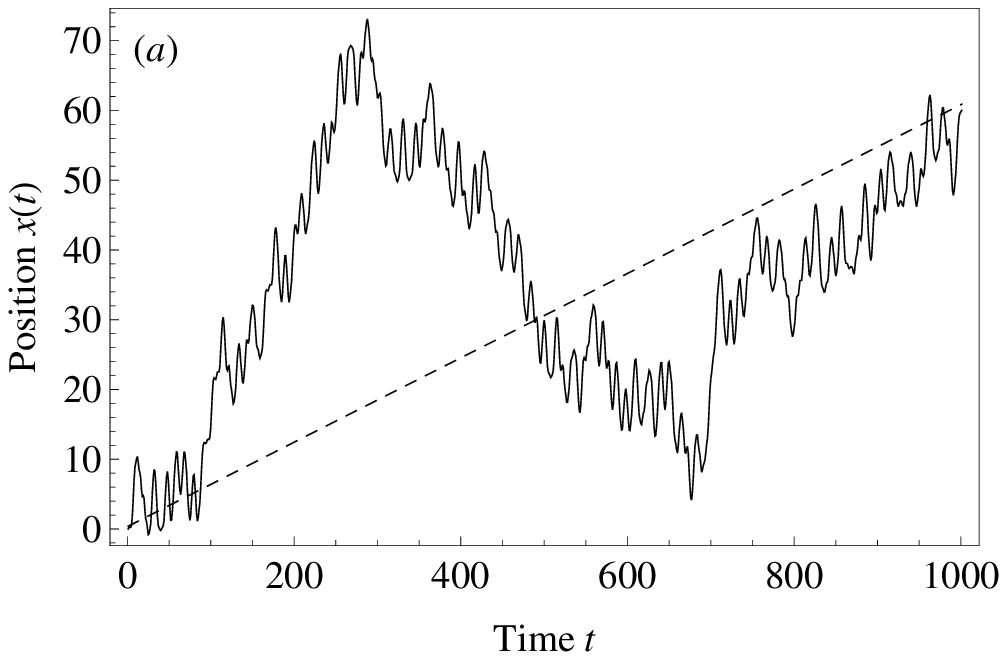}
\includegraphics[width=230pt]{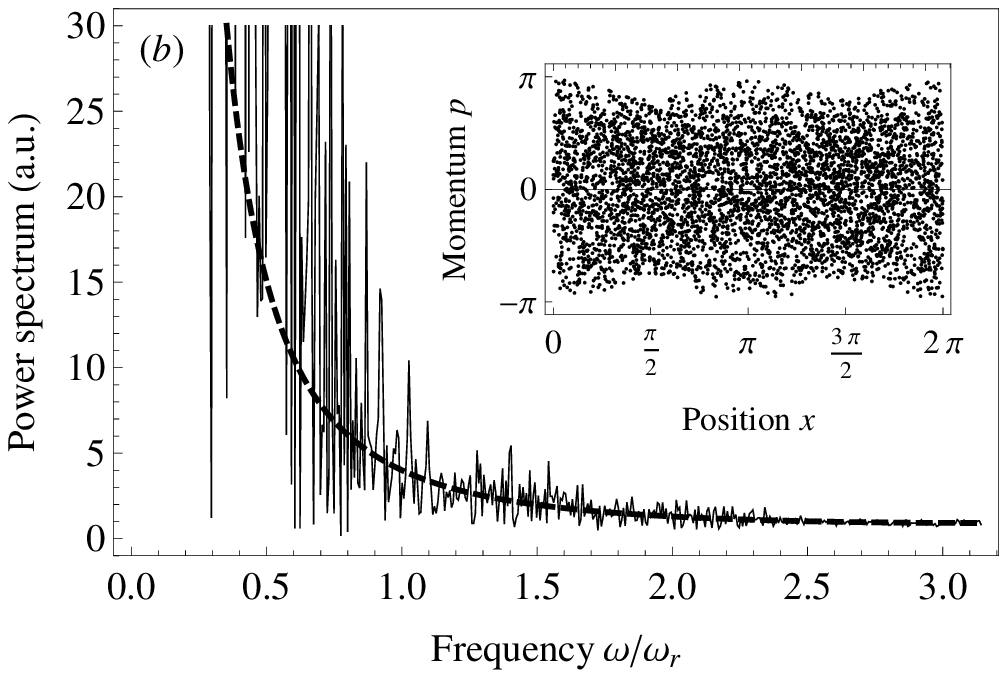}
\end{center}
 \caption{(a) A typical random walking of a cold atom in
varying standing field of $\omega=0.8$, $\lambda =2$ with $x(0)=0$,
$p(0)=0.4$. The dashed line is the average motion of the atom. (b)
The corresponding power spectrum of the random walking (solid) and
the average motion (dashed). The inset picture is the pase-space
distribution of the walking mapped to $[0,2\pi]$.} \label{figure5}
\end{figure}
In this case, the atom will exhibit a random like walking that is
shown in Fig.\ref{figure5}(a) by a typical sample of moving sequence
along optical lattice, displaying an unpredictable position of the
atom in the periodically varying lattice. This random characteristic
can be further verified by a power spectrum of this walking depicted
in Fig.\ref{figure5}(b). The dashed line in Fig.\ref{figure5}(a)
corresponds to the effective uniform motion of this walking with an
average momentum of $\langle p(t)\rangle_T\approx 0.4+0.06$ ($T$ is
the total walking time), whose fourier spectrum is depicted by a
thick dashed line for reference. The inset picture of
Fig.\ref{figure5}(b) is a long time pase-space distribution of the
atom by mapping its walking position into the first periodic region
of the lattice $[0,2\pi]$. The uniform phase-space distribution is a
statistical proof of the random properties coming up in this
deterministic system where regular and chaotic regions coexist in
this walking process.
\begin{figure}[htp]
\begin{center}
\includegraphics[width=230pt]{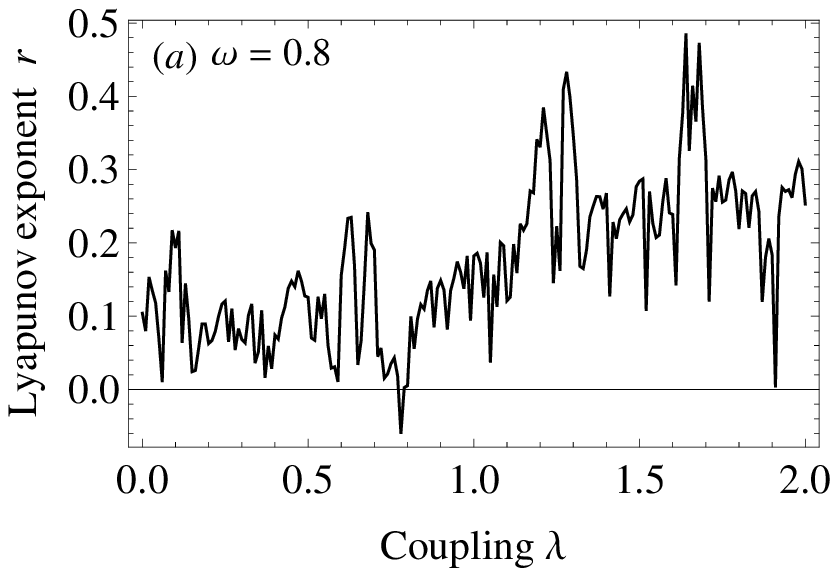}
\includegraphics[width=230pt]{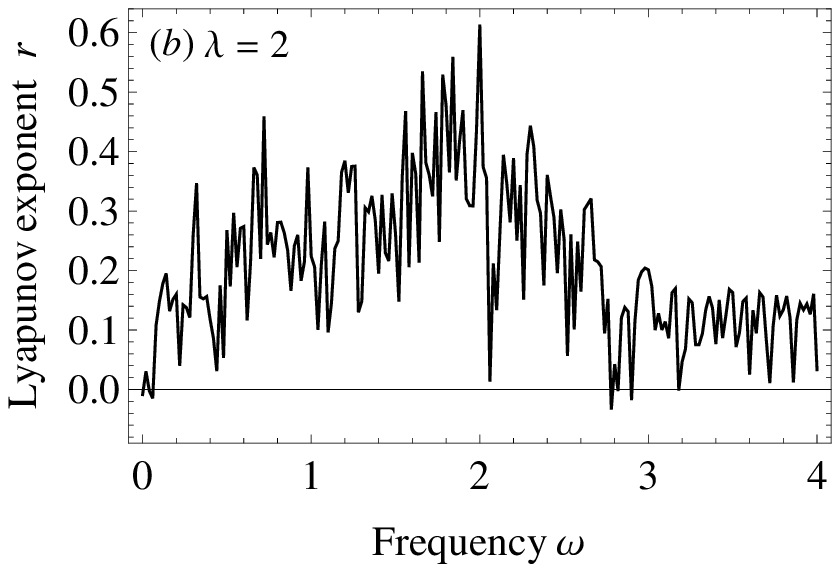}
\end{center}
\caption{(a) The Lyapunov exponent of the atomic walking versus
coupling intensity $\lambda$ with $\omega=0.8$. (b) The Lyapunov
exponent versus modulation frequency $\omega$ with $\lambda=2$. The
initial atomic position $x(0)=0$, momentum $p(0)=0.4$.}
\label{figure6}
\end{figure}

In the sense of walking sensitivity to the atomic state, above
behavior can be treated as a chaotic transportation in the standing
wave lattice similar to the results in \cite{Prants1,Prants}. The
estimation of Lyapunov exponents $\lambda_i$ can give a rough
analysis of this chaotic property. This system is equivalent to a
3-dimensional autonomous system with one Lyapunov exponent being
zero (the third dimension is $\theta=\omega t$). Because the sum of
the exponents for a Hamiltonian system is null, i.e., $\sum_i
\lambda_i=0$, therefore the Lyapunov exponents of this system must
be $-r,0,+r$. Fig.\ref{figure6} gives a rough estimate of the
largest Lyapunov exponent $r$ changing with the trapping depth
$\lambda$ (Fig.\ref{figure6}(a)) and with the modulation frequency
$\omega$ (Fig.\ref{figure6}(b)), indicating a weak chaotic
transportation along the standing wave lattice. The chaotic
transportation of this system under the influence of internal
transition has been investigated by Argonov and Prants
\cite{Prants1,Prants2,Prants}, but our results reveal that only the
external walking of an atom in a slowly varying optical lattice can
present the similar chaotic behavior. Certainly, the similar fractal
tunneling time \cite{Prants3} through a lattice with a certain
length can also be found in this walking.

\subsection{Quasi-periodic trapping state and its stability}

\begin{figure}[tbp]
\begin{center}
\includegraphics[width=220pt]{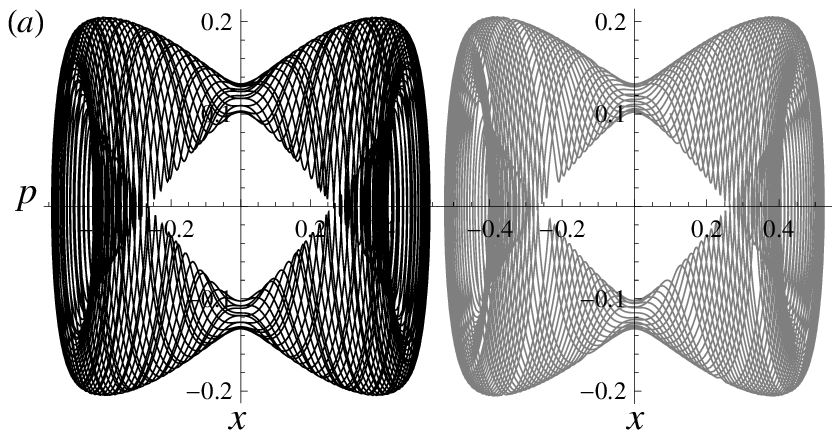}
\includegraphics[width=200pt]{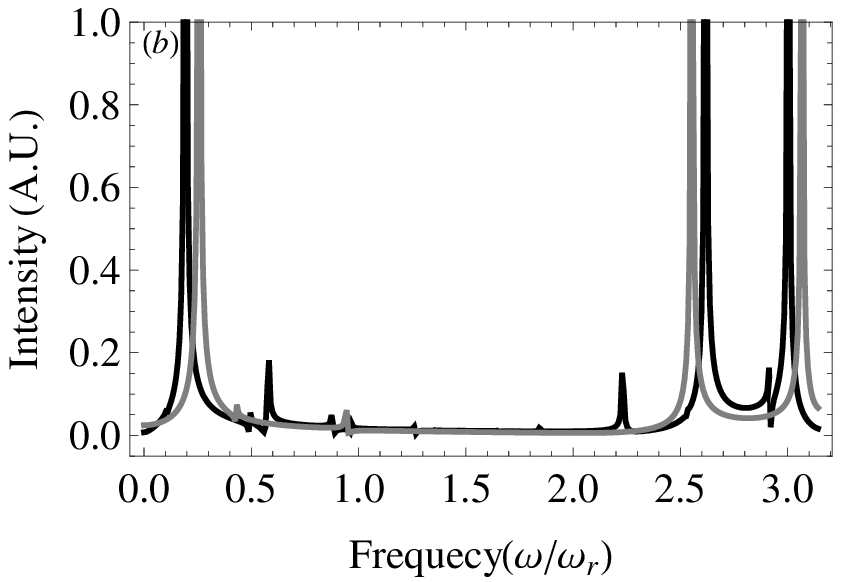}
\end{center}
\caption{(a)The phase portrait of strict solution of
Eq.(\protect\ref{Dyeqution1}) (left black orbit) and the Mathieu
solution of Eq.(\protect\ref{Msol}) (right gray orbit) with
parameters $\protect\lambda =1,\protect\omega =2.8$ and the initial
position and momentum being $x(0)=0.2$, $p(0)=0.18$. (b)The
corresponding power spectrum of the atomic momentum for the strict
solution (black line) and the Mathieu solution (gray line).}
\label{figure7}
\end{figure}

In some parametric regions the atom will be totally trapped by the
quickly varying standing wave. Fig.\ref{figure7}(a) gives an example
of this case when the frequency reaches to $\omega=2.8 \omega_r$
(left black bowknot orbit). In this case the atom takes a periodic
or quasi-periodic oscillation around a stable site near initial
position $x(0)$. In the following, we will give some detail analysis
about this trapped walking behavior.

\begin{figure}[htp]
\begin{center}
\includegraphics[width=230pt]{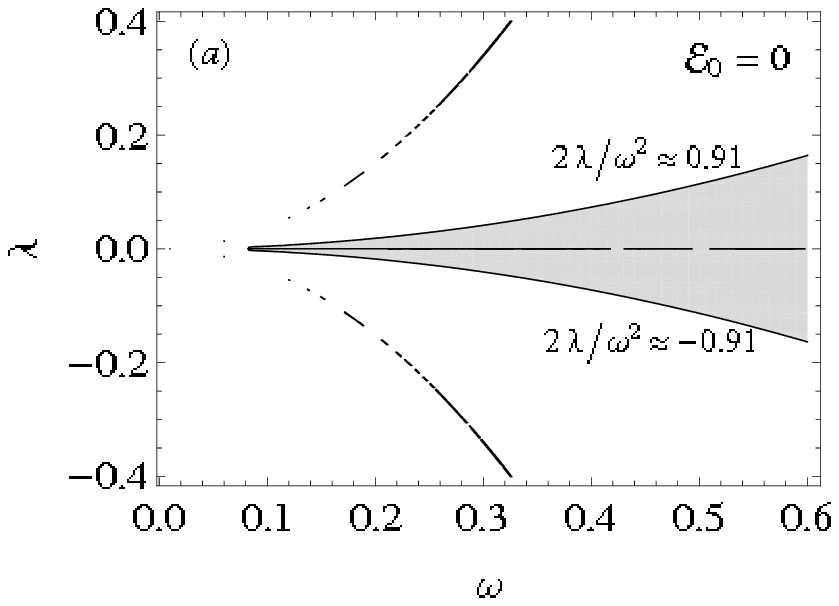}
\includegraphics[width=230pt]{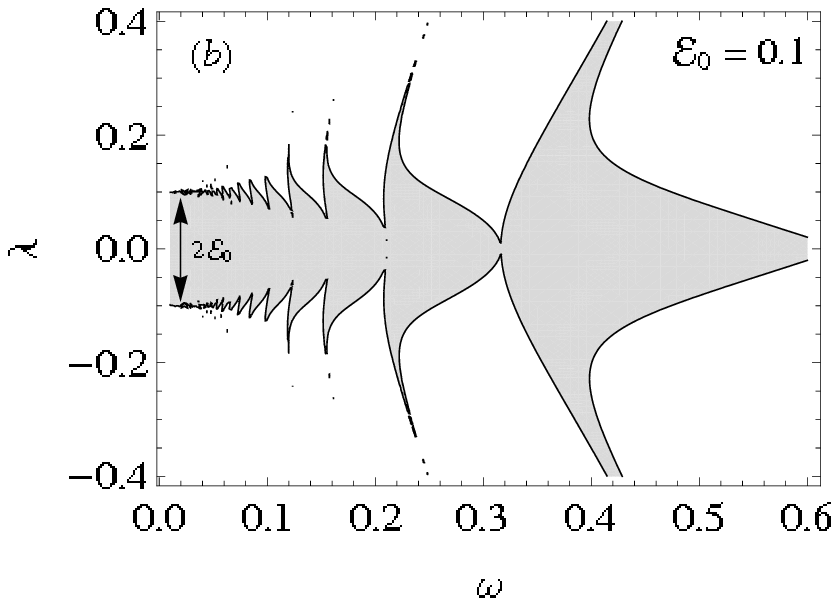}
\end{center}
\caption{The stability region for trapping state of linear solutions
Eq.(\ref{Masol}) in the $\lambda-\omega$ plane (a) for
$\mathscr{E}_{0}=0$ and (b) for $\mathscr{E}_{0}=0.1$ The shaded
area stands for the stable region.} \label{figure8}
\end{figure}

For an arbitrary varying frequency $\omega $, Eq.(\ref{Dyeqution1})
corresponds to a pendulum with changing length. The normal method of
solving this nonlinear differential equation is the linearization
approach. For the trapped atom that is very near to a stable antinode
of $\sin x^{*}=0$, for example, a small $x$ near zero is $ \sin
x\sim x$, Eq.(\ref{Dyeqution1}) will be reduced to
\begin{equation}
\ddot{x}+\left( \lambda \cos \omega t\right) x=0,\label{III-1}
\end{equation}%
which is a linear equation with a periodic coefficient and can be
solved by Floquet's theorem \cite{Teschl}. The linearized equation
(\ref{III-1}) is a Mathieu equation
\begin{equation*}
\frac{d^{2}x}{dt^{2}}+\left[ a-2q\cos \left( 2t\right) \right] x=0
\label{Mathieu}
\end{equation*}%
with a standard form of solution
\begin{equation*}
x(t)=e^{i \mu t}X(t),
\end{equation*}%
where $\mu$, in general, is a complex function of $a$ and $q$ called
characteristic exponent, and $X(t)$ is a periodic function.
Explicitly, the solution of equation (\ref{III-1}) reads%
\begin{equation}
x\left( t\right) =x(0)C\left( 0,-\frac{2\lambda }{\omega
^{2}},\frac{\omega t}{2}\right) +p(0)S\left( 0,-\frac{2\lambda
}{\omega ^{2}},\frac{\omega t}{2}\right),\label{Msol}
\end{equation}%
where $C\left( 0,-\frac{2\lambda }{\omega ^{2}},\frac{\omega
t}{2}\right)$ and $S\left( 0,-\frac{2\lambda }{\omega
^{2}},\frac{\omega t}{2}\right)$ are Mathieu even and odd functions.
Above solution can be used to estimate a variety of behavior of
Eq.(\ref{Dyeqution1}) under the trapping condition (for that the
linearization condition is valid). Fig.\ref{figure7}(a) is a
comparison of Methieu solution (right gray orbit) with the strict
simulation (left black orbit) in phase space when Methieu function
is in its stable parametric region. The figures show a nice match of
the linear approximation with Eq.(\ref{Dyeqution1}) and the atom
conducts a quasi-periodic oscillation around $x^{*}=0,\pm\pi,\pm
2\pi,\cdots$. The quasi-periodic properties of this walking can be
verified by the power spectrum of atomic momentum in
Fig.\ref{figure7}(b), where several frequency components are
manifest but the atom never exactly repeat its walking orbit in the
phase space. Certainly, on some parameters, the periodic walking is
also available. This can be discussed by the characteristic exponent
of Mathieu function. The characteristic exponent,
$\mu(a,q)=\mu(0,-2\lambda /\omega ^{2})$, predicts that only for
certain value of $\lambda $ and $\omega $ can the solution be
periodic. For parametric values of $2\lambda/\omega^2>0.91$, $\mu$
will be a complex number and Eq.(\ref{III-1}) is unstable (see the
borderlines $2\lambda/\omega^2\approx \pm 0.91$ shown in
Fig.\ref{figure8}(a)). Therefore we can determine the stable or
unstable parametric region of the linearization equation under
independent initial conditions. However in the real system, the
modulating amplitude will include a small constant part
$\mathscr{E}_0$ \cite{Mouchet} that is governed by
\begin{equation*}
\frac{d^{2}x}{dt^{2}}+\left( \mathscr{E}_{0}+\lambda \cos \omega
t\right) x=0,
\end{equation*}%
and this enables a normal Mathieu solution of
\begin{equation}
x\left( t\right) =x(0)C\left( \frac{4\mathscr{E}_{0}}{\omega
^{2}},-\frac{2\lambda }{ \omega ^{2}},\frac{\omega t}{2}\right)
+p(0)S\left( \frac{4\mathscr{E}_{0}}{\omega ^{2}} ,-\frac{2\lambda
}{\omega ^{2}},\frac{\omega t}{2}\right) .\label{Masol}
\end{equation}

Fig.\ref{figure8} gives a stability diagram of solution
Eq.(\ref{Masol}) near stable positions $x^{*}$ with the shaded area
indicating stable region of the trapping states. Fig.\ref{figure8}
shows an irregular stable parametric region of the walking
determined by the lattice depth $\lambda$ and the modulation
frequency $\omega$, with a symmetric structure for $\lambda$ and a
fractal boundary for small $\omega$. Fig.\ref{figure8}(b) also
indicates that although the symmetric gap (indicated by the arrows
in Fig.\ref{figure8}(b)) of the stable region, $2\mathscr{E}_{0}$,
in small frequency $\omega$ region goes to zero when
$\mathscr{E}_{0}\rightarrow 0$, the complicated structure of stable
region will not disappear in Fig.\ref{figure8}(a) for a real
walking. This result reveals that the dynamic stability of atomic
walking around antinodes in a varying lattice is very sensitive to
the fluctuation of the optical field \cite{Steck2}. Actually the
fluctuation of the lattice field will definitely make the atomic
walking complicated in the small modulation frequency region.
Strangely, when modulation frequency is small, the atomic walking
around antinodes can stay unstable no matter how large the depth of
the lattice is. When the frequency becomes higher, the stable region
will expand and the walking is more preferred to a stable motion.
According to Floquet's theory, the characteristic exponent $\mu$ of
the trapping solution will be $\lim_{\omega\rightarrow
\infty}\mu(0,-2\lambda /\omega ^{2})=0$, which indicates a free
atomic walking limit in a quickly varying lattice.

We should note here that the stable region of the real system
Eq.(\ref{Dyeqution1}) is not only vulnerable to the control
parameters $\lambda, \omega$ but also heavily depends on the dynamic
itself, i.e., the atom's position and momentum are also key
parameters to determine the stability of atomic walking
\cite{Teschl}. The nonlinear position-dependent of walking in
Eq.(\ref{Dyeqution1}) will introduce an effective part of
$\mathscr{E}_{0}$ to shift the atom along the lattice, which
actually breaks the spatial symmetry of the lattice walking.
Naturally, an increase of atomic momentum will definitely reduce the
stability region of the real trapping solution because the heated
atom will more easily escape from the trapping region. The stable
region of the linear solution indicates that the atom can surely be
trapped in a shallow lattice field (small $\lambda$) with a higher
modulation frequency $\omega$.

\subsection{Free ballistic flying}

\begin{figure}[tbp]
\begin{center}
\includegraphics[width=215pt]{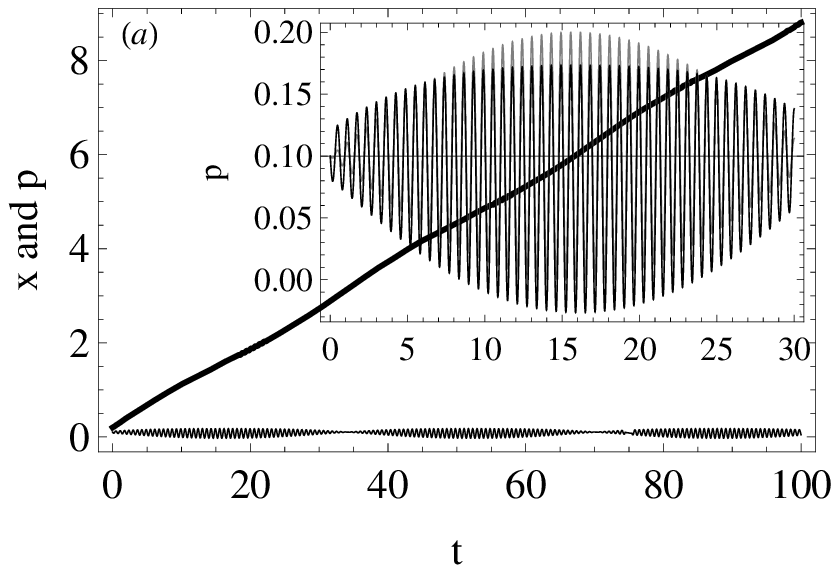}
\includegraphics[width=220pt]{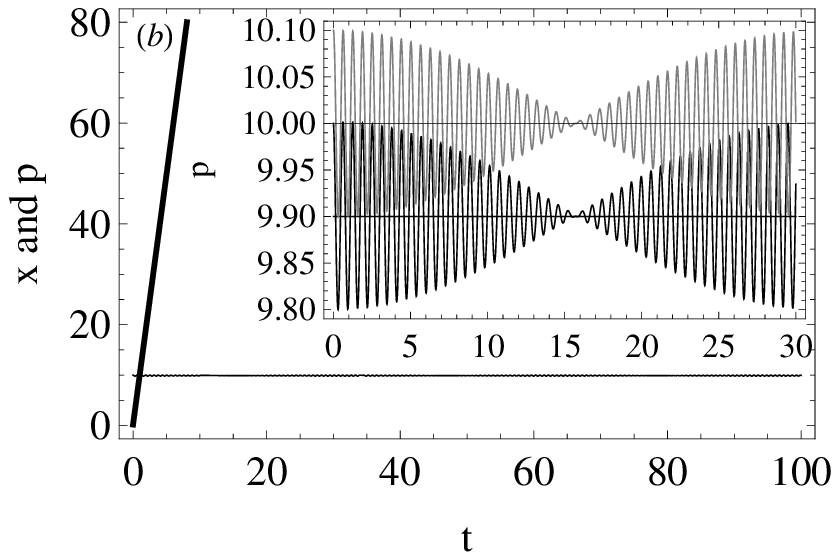}
\end{center}
\caption{The free flying of the atom (a) in a quickly varying field
with $\omega=10$, $p(0)=0.1$, $x(0)=0.2$, $\lambda =1$ and (b) in a
high speed under parameters $p(0)=10$, $\omega=0.1$, $x(0)=0.2$,
$\lambda =1$. The insets are atomic momenta zoomed in between a time
interval of $[0,30]$ with the black lines for the strict equation
and the grey ones for Eq.(\ref{beat}).} \label{figure9}
\end{figure}

Finally, we consider two extreme conditions that the modulation
frequency of the lattice is high ($\omega \gg \sqrt{\lambda}$) and
the velocity of the atom is quick ($p \gg 2 \sqrt{\lambda}$). For a
large modulating frequency $\omega$, the effective coupling constant
meets $\lambda \cos(\omega t)$ $\rightarrow 0$ and the equation can
be simplified by
\begin{eqnarray}
\ddot{x} \approx 0, \quad \omega \gg 1, \label{ballistic}
\end{eqnarray}%
which means a negligible influence of light on atomic walking and
implies a free flying of the atom in optical lattice, i.e., $x\left(
t\right) \approx x\left( 0\right) +p\left( 0\right) t$ and $p\left(
t\right) \approx p\left( 0\right)$. In the other hand, when the
speed of the atom is high, the atom also feels a quick oscillation
of the force with a frequency of $p+\omega$ in one direction and
$p-\omega$ in the opposite (see Eq.(\ref{force})). Both above cases
are displayed in Fig.\ref{figure9}(a) and Fig.\ref{figure9}(b) by
direct simulations of Eq.(\ref{Dyeqution1}) where the thick black
line, $x(t)$, indicate a nearly free flying motion and its slope is
roughly determined by the initial momentum $p_0\equiv p(0)$, showing
a small relative change of atomic momentum during the walking
process. The enlarged figures of the momenta between time interval
$[0,30]$ shown by the insets of Fig.\ref{figure9} demonstrate a beat
oscillation of the atomic momentum around the initial value, with
the average momentum satisfying $\langle p\rangle\approx p_0$.
Therefore, the oscillation behavior can be approximately described
by Eq.(\ref{force}),
\begin{equation*}
\frac{dp}{dt}=-\lambda \sin \left( p_0 t\right) \cos \left( \omega
t\right),
\end{equation*}
and an integral of it gives
\begin{equation}
p(t)=p_{0}+\frac{\lambda }{2}\frac{\cos \left[ (p_{0}-\omega )t\right] }{%
p_{0}-\omega }+\frac{\lambda }{2}\frac{\cos \left[ (p_{0}+\omega )t\right] }{%
p_{0}+\omega },\label{beat}
\end{equation}
which clearly reveals a beat oscillation of the momentum.
Comparisons of solution Eq.(\ref{beat}) (the gray lines of the
insets) with the strict solution (the black lines of the insets) are
depicted in Fig.\ref{figure9}(a) and Fig.\ref{figure9}(b). The
simulation indicates a good agreement of Eq.(\ref{beat}) with the
strict Eq.(\ref{Dyeqution1}) except for some minor differences.
Besides, Eq.(\ref{beat}) also suggests a resonance behavior of the
momentum when atomic momentum meets $p_0\approx\pm\omega$. Under
resonant condition, Eq.(\ref{beat}) becomes a bad approximation
because the momentum will be divergent. However, as long as the
condition of $\omega \gg \sqrt{\lambda}$ or $p_0 \gg
2\sqrt{\lambda}$ is satisfied, the momentum enhancement by the
resonant effect in real system remains small relative to the large
initial momentum, and the atom still keeps its ballistic flying
under resonance condition. However, if the initial atomic momentum
is small, this resonance effect will be manifest only under small
modulation frequencies (see Eq.(\ref{beat})) and this resonant
behavior will also be suppressed by the nonlinear character of a
real walking. Therefore, in conclusion, if the modulation frequency
is high and the atomic momentum is large, the atom will almost
conduct a free ballistic flying with its momentum taking a beat
oscillation roughly around its initial value.

\section{Conclusions}
\label{sec4}%
A single atom coupled with high-finesse cavities is a fundamental
system for quantum research, and many theoretical and experimental
works have been done on this system
\cite{Teper,Stibor,Ye,Kuhr,Wilk,Rempe,Kimble,Ritsch}. Directly based
on quantum description, the quantum aspect of the dynamics are
usually focused on, such as on nonclassical statistics,
entanglement-induced effects, quantum information processing, which,
contrarily, makes the classical contribution obscure. While some
authors on this system concern more about the transmission of the
light through a cavity to detect or control the atomic dynamics
\cite{Rempe}, such as detecting atomic trapping states, or
controlling atomic trajectories \cite{Kimble1,Kimble2}. However all
the atomic motion considered above are closely involved with the
internal dynamic and the external dynamics is still mixing in atomic
control. In order to investigate the external dynamic in a classical
point of view, the internal variables should be decoupled or traced
out. Raizen and his coworkers \cite{Raizen} resort to a dynamical
map (periodic kicked rotator) based on pulsed standing field to
investigate the atomic dynamic behavior and find a good agreement
with the classical dynamic in a noisy environment. Particularly,
their experiment studies on the motion of cold cesium atoms in an
amplitude-modulated standing wave of light \cite{Steck1,Steck2} have
a very close relation to our study in this paper. However, we use a
simple classical model under a resonant condition and find a more
rich dynamic behaviors of one atom in the optical lattice beyond a
direct influence of internal dynamics.

As the fluctuating environment of the cavity mode, trapping a single
atom in the cavity for a long time is turned out to be difficult.
Therefore, sensitivity of the atomic motion to field modulation is a
key problem for a practical one-atom control \cite{Steck2}. In this
paper, we closely consider the classical walking of an atom in an
field-controlled standing wave and reveals a diverse dynamic region
of atomic motion. In the parametric region of lower modulation
frequency, an oscillation with classical random jumping is found for
cold atom. With the increasing of frequency, random atomic motion,
chaotic transportation and the quasi-periodic trapping state appear.
If the modulation frequency is high or the velocity of the atom in
the lattice is large, the atom will exhibits a ballistic fly with a
momentum beating oscillation. The study of dynamic stability shows
that the transition between these dynamical regimes is irregularly
determined not only by the lattice field but also by the position
and momentum of the atom. Our results indicates a wide parametric
region of unstable walking and a susceptibility of stable motion to
the field fluctuations of amplitude and frequency as well as to the
motion itself. Therefore, this work gives a rich clue for atomic
control in the cavity and also provides a useful insight into the
dynamical behavior of atoms in a periodically varying lattice.

\section*{Acknowledgements}
This work is supported by the
Shaanxi Provincial Natural Science Foundation (No.SJ08A12) and the
National Science Foundation Project (No.10875076).
%
\appendix

\section{The derivation of Eq.(\ref{DRM-1})}
The dynamic of atom in a\ certain field can be manipulated by
different configuration of $\mathbf{E}(\mathbf{r},t)$ according to
the Hamiltonian of Eq.(\ref{H}). A standing wave, generated by an
interference of two quasi-monochromatic field with an envelop,
$\mathscr{E}\left( \mathbf{r},t\right) $, and the polarization,
$\mathbf{e}_{\lambda }$, propagating in opposite direction, will be
\begin{eqnarray*}
\mathbf{E}(\mathbf{r},t) &=&\frac{1}{2}\mathbf{e}_{\lambda
}\mathscr{E} \left( \mathbf{r},t\right) \left[ \cos \left( \nu
t+\mathbf{k}\cdot \mathbf{r}\right) +\cos \left( \nu t-\mathbf{k}
\cdot \mathbf{r}\right) \right] \\
&=&\mathbf{e}_{\lambda }\mathscr{E}\left( \mathbf{r},t\right) \cos
\left( \mathbf{k}\cdot \mathbf{r}\right) \cos \left( \nu t\right) ,
\end{eqnarray*}%
where $\nu $ is the carrier frequency and the slowly varying
temporal envelop, $\mathscr{E}\left( \mathbf{r},t\right) $, can be
controlled in the experiment by changing the amplitude of the input
field. If we stabilize the phase of the field in a linear cavity
along $x$ axis, the Hamiltonian of Eq.(\ref{III-2}) will be got.
More generally, we can write down the Schr\"{o}dinger equation in
the standing wave as
\begin{equation*}
i\hbar \frac{\partial }{\partial t}\Psi \left( \mathbf{r},t\right)
=\left[ \frac{\hat{p}^{2}}{2m}+\hat{H}_{a}-\Omega \left(
\mathbf{r},t\right) \cos \left( \nu t\right) \right] \psi \left(
\mathbf{r},t\right) ,
\end{equation*}%
where
\begin{equation*}
\Omega \left( \mathbf{r},t\right) =\mathbf{\hat{d}}\cdot
\mathbf{e}_{\lambda }\mathscr{E}\left( \mathbf{r},t\right) \cos
\left( \mathbf{k}\cdot \mathbf{r} \right),
\end{equation*}%
and $\hat{H}_{a}$ denotes the energy of the internal state: on
ground state $\left\vert 1\right\rangle$ with energy $\epsilon
_{1}$, on excited state $ \left\vert 2\right\rangle$ with energy
$\epsilon _{2}$. In real space, the atomic wavefunction can be
written by a two-component form of
\begin{equation*}
\Psi \left( \mathbf{r},t\right) =\varphi _{1}\left(
\mathbf{r},t\right) e^{-i\epsilon _{1}t/\hbar }\left\vert
1\right\rangle +\varphi _{2}\left( \mathbf{r},t\right) e^{-i\epsilon
_{2}t/\hbar }\left\vert 2\right\rangle.
\end{equation*}%
Under the rotating wave approximation, the Schr\"{o}dinger equation
for the two-component wave function is
\begin{equation*}
\frac{\partial }{\partial t}\left(
\begin{array}{c}
\varphi _{1} \\
\varphi _{2}
\end{array}%
\right) =i\left(
\begin{array}{cc}
\frac{\hbar }{2m}\nabla ^{2} & \Omega _{R}\left( \mathbf{r},t\right)
e^{-i\delta t} \\
\Omega _{R}\left( \mathbf{r},t\right) e^{i\delta t} & \frac{\hbar
}{2m} \nabla ^{2}
\end{array}%
\right) \left(
\begin{array}{c}
\varphi _{1} \\
\varphi _{2}
\end{array}%
\right)
\end{equation*}%
where the detuning frequency $\delta =\omega _{0}-\nu $ and the
transition frequency $\omega _{0}=\left( \epsilon _{2}-\epsilon
_{1}\right) /\hbar $. The Rabi frequency is defined by
\begin{equation*}
\Omega _{R}\left( \mathbf{r},t\right) =\frac{\Omega \left(
\mathbf{r} ,t\right) }{2\hbar }=\frac{\mathbf{\hat{d}}\cdot
\mathbf{e}_{\lambda }}{2\hbar }\mathscr{E}\left( \mathbf{r},t\right)
\cos \left( \mathbf{k}\cdot \mathbf{r}\right).
\end{equation*}

In the experiment, the carrier frequency $\nu $ is adjusted to the
transition frequency $\omega _{0}$ (resonance $\delta =0$), then%
\begin{equation*}
i\hbar \frac{\partial }{\partial t}\left(
\begin{array}{c}
\varphi _{1} \\
\varphi _{2}%
\end{array}%
\right) =\left(
\begin{array}{cc}
-\frac{\hbar ^{2}}{2m}\nabla ^{2} & -\hbar \Omega _{R}\left(
\mathbf{r},t\right)  \\
-\hbar \Omega _{R}\left( \mathbf{r},t\right) & -\frac{\hbar
^{2}}{2m}\nabla ^{2}
\end{array}%
\right) \left(
\begin{array}{c}
\varphi _{1} \\
\varphi _{2}
\end{array}%
\right)
\end{equation*}%
In this case above equation can be decoupled by introducing \cite{Cook}
\begin{eqnarray*}
\psi _{+}\left( \mathbf{r},t\right) &=&\frac{1}{\sqrt{2}}\left[
\varphi _{1}\left( \mathbf{r},t\right)
+\varphi_{2}\left(\mathbf{r},t\right) \right] , \quad \psi
_{-}\left( \mathbf{r},t\right) =\frac{1}{\sqrt{2}}\left[ \varphi
_{1}\left( \mathbf{r},t\right) -\varphi
_{2}\left(\mathbf{r},t\right)\right] ,
\end{eqnarray*}%
with equations of
\begin{eqnarray*}
i\hbar \frac{\partial }{\partial t}\psi _{+}
&=&-\frac{\hbar^{2}}{2m}\nabla^{2}\psi _{+} -\hbar \Omega _{R}\left(
\mathbf{r},t\right) \psi _{+}, \quad
i\hbar \frac{\partial
}{\partial t}\psi _{-} =-\frac{\hbar^{2}}{2m}\nabla ^{2}\psi
_{-}+\hbar \Omega _{R}\left(\mathbf{r},t\right) \psi _{-}.
\end{eqnarray*}%
Above equation for wave functions $\psi _{\pm }\left(
\mathbf{r},t\right) $ indicates that the atom on any state will feel
two different potentials
\begin{equation*}
V_{\pm }\left( \mathbf{r},t\right) =\mp \hbar \Omega _{R}\left(
\mathbf{r} ,t\right) =\mp \frac{\mathbf{\hat{d}}\cdot
\mathbf{e}_{\lambda }}{2} \mathscr{E}\left( \mathbf{r},t\right) \cos
\left( \mathbf{k}\cdot \mathbf{r} \right) .
\end{equation*}%
If the envelop function of the field is controlled by (amplitude
modulation field)
\begin{equation*}
\mathscr{E}\left( \mathbf{r},t\right) =g_{0}\cos \left( \chi
t\right) ,
\end{equation*}%
the effective potential will be%
\begin{eqnarray*}
V\left( \mathbf{r},t\right)  &=&\mp g_{0}\frac{\mathbf{\hat{d}}\cdot
\mathbf{e} _{\lambda }}{2}\cos \left( \chi t\right) \cos \left(
\mathbf{k}\cdot \mathbf{r}\right) \equiv \hbar \lambda \cos \left(
\chi t\right) \cos \left( \mathbf{k}\cdot \mathbf{r}\right) .
\end{eqnarray*}%
For a linear cavity, the effective potential we pick up to consider
is
\begin{equation*}
V\left( x,t\right) =\hbar \lambda \cos \left( \chi t\right) \cos
\left( kx\right) ,
\end{equation*}%
where $\lambda $ is the effective coupling parameter and $\chi $ is
the modulation frequency of the field amplitude.
%


\begin{thebibliography}{00}
\bibitem{1} In a real atomic-scale experiment, one atom can be traced and detected with a high resolution beyond the Heisenberg uncertainty principle.
\end{thebibliography}
\end{document}